# EXPERIMENTAL APPROACH FOR OPTIMIZING DRY FABRIC FORMABILITY

S. Allaoui[1*], G. Hivet[1], A. Wendling[1], D. Soulat[1], S. Chatel[2]

[1]Institut PRISME/MMH, Université d'Orléans, 8 rue Léonard de Vinci 45072 Orléans Cedex2, France
[2] EADS France, Innovation Works, Mechanical Modelling Research Team, 12 rue Pasteur,BP76, 92152 Suresnes Cedex, France
*samir.allaoui@univ-orleans.fr;

**Abstract**

*In order to understand the mechanisms involved in the forming step of LCM processes and provide validation data to numerical models, a specific experimental device has been designed in collaboration between PRISME Institute and EADS. This toot also makes it possible to test the feasibility to obtain specific double curved shape constituted with dry fabric reinforcement. It contains one mechanical module containing the classical tools in forming process, (punch, blank holder, and open-die), and one optical module to measure the 3D-deformed shape and the distribution of local deformations, like shear angles of the woven reinforcement during all the process. The goal of this paper is to present the potentialities and the first results obtained with this device.*

## 1. Introduction

Among the manufacturing processes, Liquid Composites Moulding are undoubtedly among the most interesting for the obtaining of a composite part. It offers several advantages like reduced solvent emissions, part quality, and process repeatability [1]. Furthermore, forming processes have shown their efficiency for the manufacturing of metallic materials for many years, so that they have replaced, in many cases, historical processes. Nevertheless, its development in industrial application is not so common today especially because the process has to be fairly improved.

LCM processes start with the forming of a dry fabric on what can be a complex (double curved) shape (figure 1). It is a delicate phase because mechanical behaviour of the final composite part depends of the reinforcement's position after forming. In addition, the angles between the warp and weft yarns also influence the permeability of the reinforcement and thus the stage of filling by the resin. This first step is not completely mastered today what constitutes a main drawback for the development of this process at the industrial level. Nowadays, the main industrial approach used to study part feasibility is trial and error.

Other approaches have to be considered in the future. Among the strategies useable to investigate a priori the formability of a given fabric on a given shape, finite element simulation and experimental demonstrators can be considered. The complementarity between these two methods will enable to understand and model accurately the performing step and will hopefully permit to decrease the cost and time needed for the tools and fabrics development.

Several numerical studies were carried out in order to optimize the forming process [2, 3, 4]. These simulations need the knowledge of the specific mechanical behaviour of fabrics, which is identified during simple experimental tests. Among these tests, we can quote: shearing test [5, 6, 7], bending test [8], friction test [9], tensile and compression test [10]. The results obtains must be validates with experimental data of woven reinforcement forming in order to validate the models and behaviour laws developed in these studies.

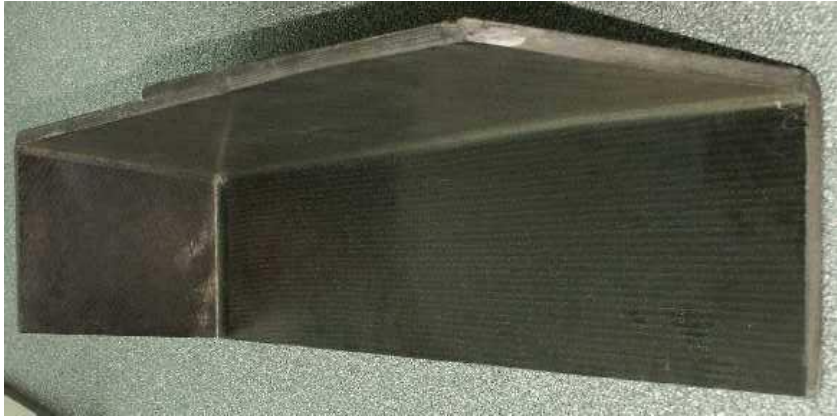

**Figure 1.** Example of double curved shape for aeronautical piece

Little study exists on the experimental validation of preforming. Hemispherical dome is the most studied perform. Mohammed and al. have proposed a mould composed by an upper and lower part [11]. The female part of the mould was made of a transparent material in order to visualize the fabric behaviour during stamping. A square grid is inscribed on the fabric with an ink pen. After performing, shear angle was identified by the comparison of the final grid to the initial one. Another tool, for the same geometry of perform, was proposed by sidhu and al. [12]. It is composed of a punch and an open die. The spherical punch, connected to a hydraulic press, makes it possible to form the hemispherical preform. Photos were taken during different stage of the draping and were used to correlate the shape of the fabric preform obtained by the numerical simulations.

This paper deals with the design of a specific demonstrator, developed with the collaboration of EADS, dedicated to the formability study of aeronautical parts and used to obtain three-dimensional double curved shape for thick parts constituted by several plies of fabric.

## 2. Presentation of the preforming device

### 2.1 Motivation

For aeronautical applications, the use of the RTM process is under study in order to obtain a three-dimensional double curved shape for thick parts constituted by several plies of fabric (figure 1) and without defects. Obviously, defects on the dry fabric preform will lead to a significant decrease of the composite performance what is completely unacceptable for flying applications.

So, the development of one specific device able to preform woven textile reinforcement has two main objectives:

- First, such a device permit to analyze experimentally the actual possibilities to manufacture any double curved composite structure without defects in the useful part with a given textile reinforcement. The role of the blank holders, of the pretensions, of the speed of the tools…can be investigated. This is very useful because the composite forming processes do not benefit from so many experiences than in the case of sheet metal forming. Furthermore the composite reinforcements are very numerous and different and it is not simple to extrapolate the results of a forming process to an other one. An example corresponding to an angle bracket used in aeronautical applications is shown Fig. 1. The design of the part using a composite material leads to a 40% weight decrease. Nevertheless the manufacturing of this part in a composite material is difficult in particular making the prefom in a RTM process with a good homogeneity of the fibre density and without defaults especially wrinkles.

- Secondly, analyze of the state of the composite reinforcement after forming is an essential mean of validating numerical simulations. The codes developed to simulate the textile reinforcement forming can significantly reduce component development time. But their validation by comparison with experimental forming processes is important for the confidence in their results.

For that purpose, a forming demonstrator should then enable to give the membranous position, orientation of the yarns, the presence of defect of any type and the 3D strains of the fabric. However these behaviours must be measured at different processing parameters: blank holder force, punch velocity, shape of the tools, fabric, number and orientation of the layers at the initial state.

## 2.2 Constitution of the device

The demonstrator is composed of three parts. The first part consists in a pair punch/open die, the die is chosen open in order to see perfectly the yarns during the stamping. The punch is moved using an electric jack in order to control easily and accurately the punch position and speed. The device is build so that any pair of punch/die can be adapted. Since our study concerns a corner of a square box, at the moment three couples have been manufactured: a square box, a tetrahedron, and a prism (figure 2) Sensors enable to track: the position, speed and load exerted by the punch.

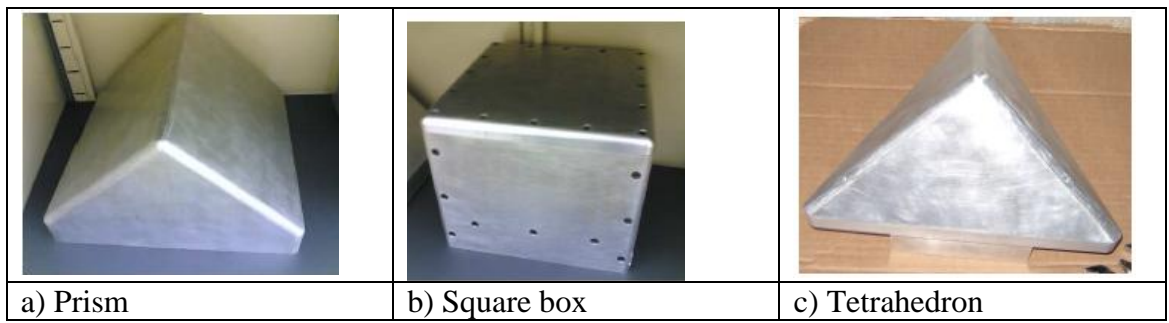

| a) Prism | b) Square box | c) Tetrahedron |

**Figure 2.** Punches

The second part of the device is constituted by a blank holder system: two mechanisms have been designed. The first one is a classical multi-part blank holder. It is composed of maximum 9 independent blank holders actuated by maximum 9 pneumatic jacks that enable to impose and sensor independently a variable pressure on each of them. Here again the system has been designed so that the contact zones of the black holders can be easily manufactured and changed to investigate the influence of their geometry on the process. The second disposal is designed to replace the blank holders by a fixed tension on the yarns extremities. This disposal is dedicated to investigate the real impact of the pressure and friction forces applied by the blank holders. It enables an easier comparison with finite element simulation since the complex phenomenon of pressure and friction between the yarns and the blank holders does not need to be modelled in this case.

The third part consists in a 3D Digital Image Correlation system. Two numerical cameras are located at the top of device, they can be positioned in function of the specific zone of the fabric that has to be analysed. For the computation of the displacement field we use marks tracking technique [13]. A speckle pattern, by additional applied paint, on the fabric surface is required (figure 3) before the test. The displacement field of these points marked on fabrics is computed with each digital image during the process and a reference image of the object. From the displacement field, the strains components are calculated in the tangential plane of the fabric, and especially the shear in plane. During the preparation of the fabric, the paint marking on yarns requires a specific attention. Points must be placed exactly on fibres in fact to track, during the process, deformation along the yarns.

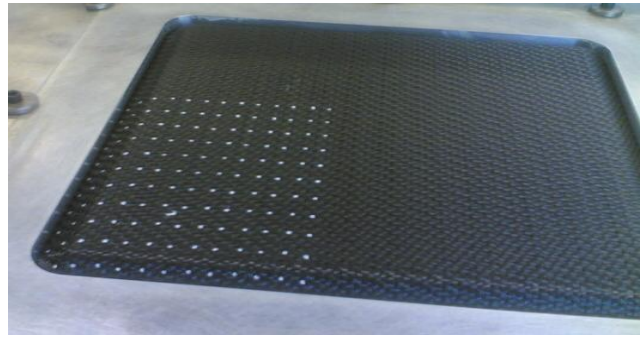

**Figure 3.** Sample with markers at the initial state

With this 3D displacement field the membranous strains can be easily computed. The thickness variation is not so easy to extract from the displacement field and is not available at the moment, but it is under study.

## 3. Experimental Results

### 3.1 Fabric used

The tests presented in this paper are carried out on a composite woven reinforcement used in aeronautics. It is denoted G1151® and constituted (figure 4) by an interlock weaving of 6K carbon yarns (630g/m², 7.5 yarns/cm).

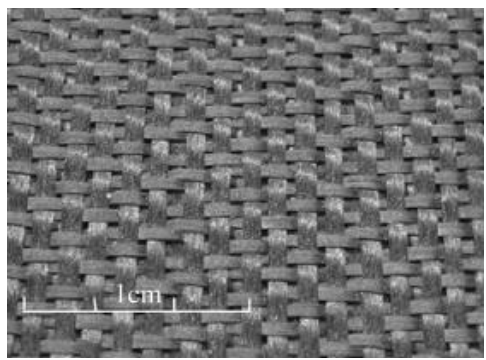

**Figure 4.** Woven carbon reinforcement (G1151®)

The main strain mode that enables fabrics to form on double curved parts is in plane shear. Thus identifying the shear behaviour of the fabric is crucial to be able to understand the forming behaviour of the fabric. The shear behaviour of the G1151® has been identified experimentally in our lab using both a specific picture frame and a bias-test. The shear curve obtained expresses the torque per area unit in function of the shear angle [14] and is presented on figure 5. This curve is a classical one as far as in pane shear of fabrics is concerned.

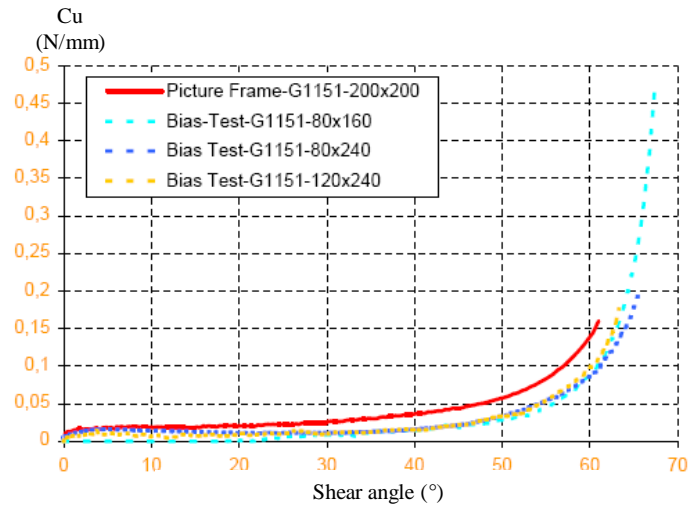

**Figure 5.** G1151® shear curve obtained using Picture Frame and Bias Test.

The shear torque is very low until the lateral contact between the yarns of the same network starts to occur. Then the shear torque increases widely and round a specific angle that is called the "locking angle", the shear stiffness becomes too high and the fabric starts wrinkling. The shear angle should then not reach the locking angle value during forming in order to avoid wrinkling.

Obviously, the wrinkling phenomena: position, shape, number…, is strongly related to the bending stiffness of the layer. The bending behaviour of the fabric has thus been investigated using a specific device developed in our lab [15].

Another way to avoid wrinkling is to introduce tension in the yarns. Increasing the blank holder pressure enables to introduce tension in the yarns due to the friction between tools and fabric. The blank holder should thus contribute to avoid wrinkling. But, since the tension will widely increase in the yarns, the biaxial behaviour of the fabric has to be identified in order to verify that the yarns are not damaged. The biaxial tensile behaviour of the G1151® has been measured on a specific biaxial tensile device developed in the lab [16, 17].

## 3.2 Defects on final preform

Two types of parameters can be distinguished which could affect the results in terms of shape, type of imperfections (like wrinkles) or in plane shear strains. The fabric parameters concern the initial form of the blank and it position, the orientation of the yarns and the number of ply. The device parameters concern the punch geometry, the number and the position of blank holder, and values of pressure, punch stroke and speed. For this study we represent on figure 4 an example for the tetrahedron punch.

The test that is presented in this paper is the forming of one layer with a tetrahedron punch. The process parameters are the following: 6 rectangular blank holders (area equal to 0.01m²), 1 bar pressure on each of them, 30mm/min punch speed, 160mm punch stroke. Different initial orientations of the fabric have been tested but only one will be presented here (figure 6): the warp direction aligned with a symmetry plane of the tetrahedron.

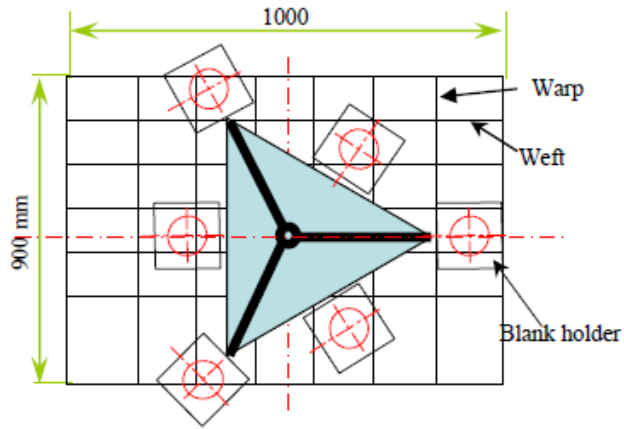

**Figure 6.** Position of the fabric.

Results are presented on figure 7. An outside view of the perform shows that the agreement with the punch shape at the end of the forming is pretty good. Wrinkles appear as it was foreseen but not in the useful part of the preform (figure 7 a).

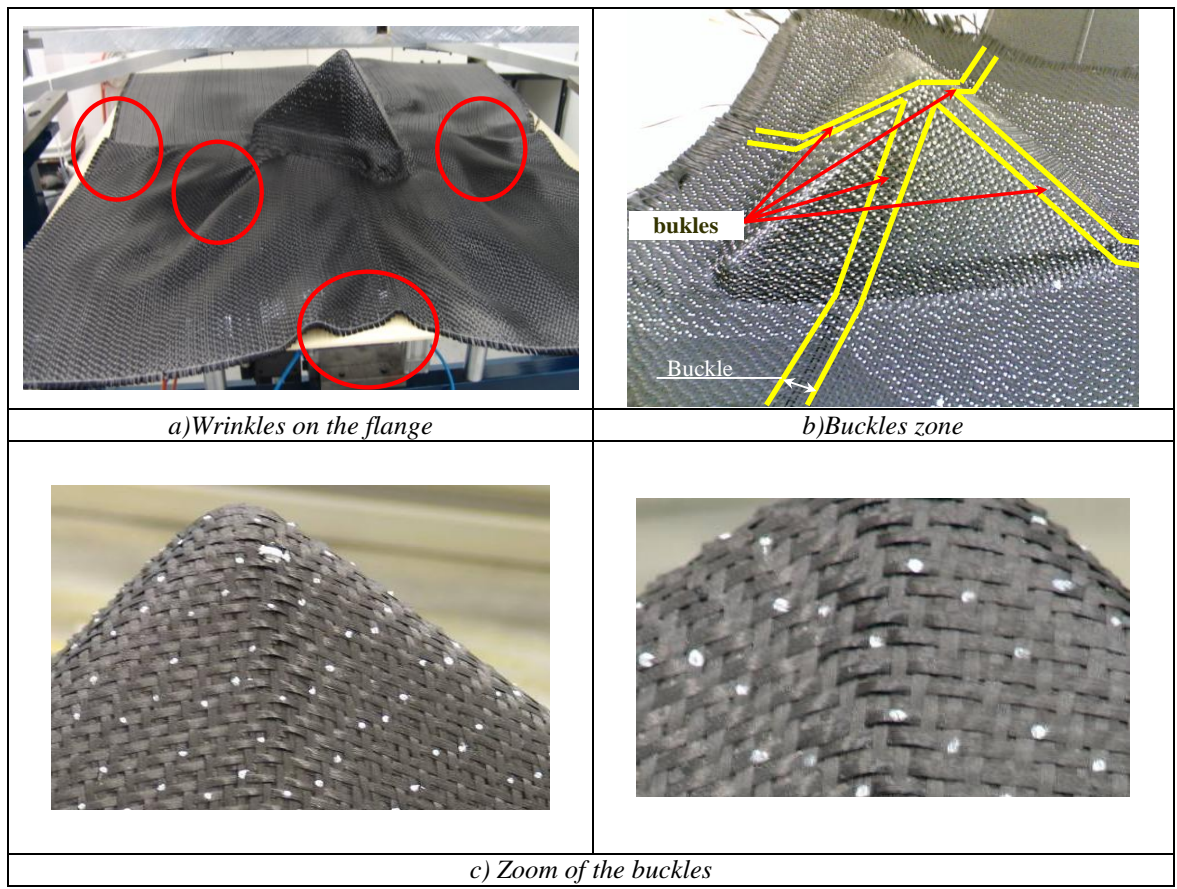

**Figure 7.** Defects at the end of the forming of a layer of G1151® with tetrahedron punch

Nevertheless a closer look on the preform enables to see another type of defect that was not predicted by the analysis of the standard membranous tests on fabrics. In four specific zones (figure 7 b, figure 7c), buckles can be observed on the transverse yarns. These 4 zones are located in small bands that follow the direction of the longitudinal yarns and pass through the tetrahedron vertex (figure 7b). It can be clearly seen (figure 7c) that transverse yarns are submitted to an out of plane bending. Their contribution of the biaxial stiffness of the membrane is then near 0, what is a real drawback for the final composites stiffness. These buckles are different from wrinkling since the membrane itself (the layer) is not wrinkled. Nevertheless, they are at least as problematic as wrinkles.

## 3.3 Optical measurement analyses

With the software of image processing it makes it possible to obtain the instantaneous position of each point marked on fabric. These positions are obtained with the three space co-ordinates. Thus, the post processing enables us to reconstruct the 3D geometry of the preform (Fig. 8).

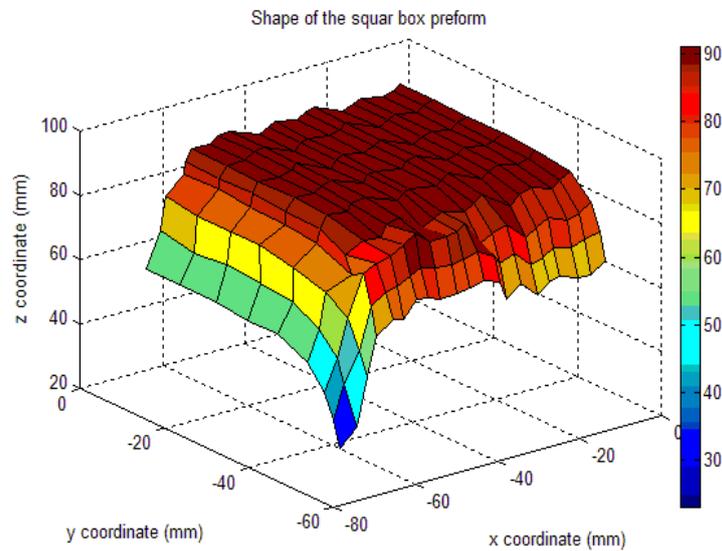

**Figure 8:** Shape of the square box preform after stamping

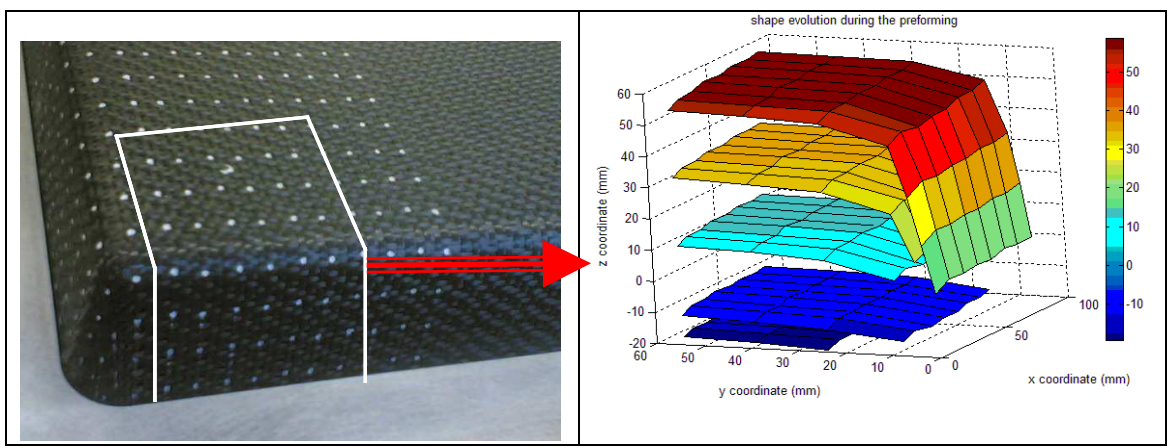

**Figure 9:** Shape evolution of an area of a square box preform during stamping

However, the data acquisition is done during all the process, enables us to analyse the behaviour of the fabric at any moment of the test, for each shape considered. Thus, figure 9 represents the evolution of the geometry shape at different stage of the stamping for an area of square box preforming. This possibility can be used to determine when defaults appears in the reinforcement or for which process parameter they appear.

The software also gives displacements, and could compute plane component of strain for the fabric. Knowing the instantaneous positions, the shear angles between yarns can thus be computed. Indeed, the local behaviour of fibres can be studied for the same test by considering only the markers along on these fibres. Figure 10 displays the measured shear angles at the end of the process on one face of the tetrahedron shape. We can conclude on a good homogeneity of this strain component. On this face, yarns are very tightening except in the buckle area, for the reasons explained before.

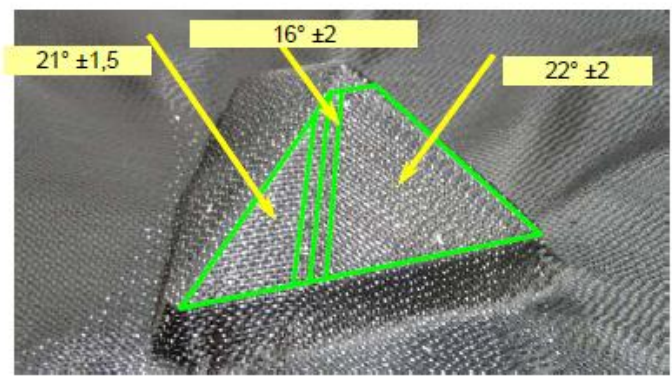

**Figure 10:** Shear angle between yarns on the tetrahedron shape

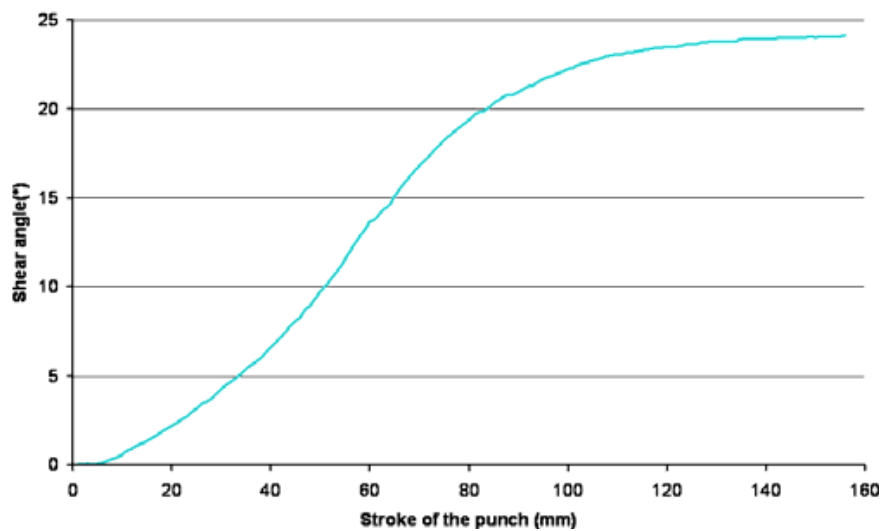

**Figure 11:** Evolution of the fabric shear angle during the process on the tetrahedron shape

The advantage of our device is to obtain the evolution of quantities not only at the end of the process but during the whole process. We represent figure 11, for example, the evolution of the shear angle in function of the punch stroke on one face of the shape. Because on this face no wrinkles are present, the shear angle does not reach the locking angle value (figure 5). We can conclude that on this face, the shear load is weak compared to the tension in yarns, and no load is opposite to the rotation between yarns.

## 4. Introduction

The specific device designed through the collaboration of EADS and the PRISME Institute presented in this paper enabled to obtain first promising results. Instrumented with digital cameras, the displacement field can be reconstructed in order to analyze the link between the defects on the perform, the yarns orientation and the mechanical behaviour of the fabric that can be identified using specific experimental disposals designed in the PRISME Institute. The large panel of process conditions can be studied in order to see their influence on the perform obtained. This tool will undoubtedly permit to increase our knowledge on the forming process of dry fabrics. A type of defect (buckles) that was not attempted has, for example, already been pointed out for the tetrahedron shape. Many works have to be done in order to achieve our goal to define a priori the optimal process parameters and fabric properties. Furthermore, besides the experimental analysis of the process, this device will also enable to provide interesting data in order to develop the numerical models.